\documentclass[sigconf]{acmart}
\AtBeginDocument{%
  }

\acmConference[Preprint]{Preprint}{2026}{}

\usepackage{subcaption}

\begin{document}

\title{RESPOND: Responsive Engagement Strategy for Predictive Orchestration and Dialogue}


\author{Meng-Chen Lee}
\orcid{0009-0002-9726-1153}
\affiliation{%
  \institution{University of Houston}
  \city{Houston}
  \state{TX}
  \country{USA}}
\email{mlee45@uh.edu}

\author{Costas Panay}
\affiliation{%
  \institution{Microsoft}
  \city{Redmond}
  \state{WA}
  \country{USA}}
\email{costas.panay@microsoft.com}

\author{Javier Hernandez}
\orcid{0000-0001-9504-5217}
\affiliation{%
  \institution{Microsoft}
  \city{Cambridge}
  \state{MA}
  \country{USA}}
\email{javierh@microsoft.com}

\author{Sean Andrist}
\orcid{0000-0003-4972-7027}
\affiliation{%
  \institution{Microsoft}
  \city{Redmond}
  \state{WA}
  \country{USA}}
\email{sandrist@microsoft.com}

\author{Dan Bohus}
\orcid{0000-0002-6283-0590}
\affiliation{%
  \institution{Microsoft}
  \city{Redmond}
  \state{WA}
  \country{USA}}
\email{dbohus@microsoft.com}

\author{Anatoly Churikov}
\affiliation{%
  \institution{Microsoft}
  \city{Redmond}
  \state{WA}
  \country{USA}}
\email{Anatoly.Churikov@microsoft.com}

\author{Andrew D. Wilson}
\orcid{0000-0001-5751-9354}
\affiliation{%
  \institution{Microsoft}
  \city{Redmond}
  \state{WA}
  \country{USA}}
\email{awilson@microsoft.com}

\renewcommand{\shortauthors}{Lee et al.}

\settopmatter{printacmref=false,printfolios=true}
\renewcommand\footnotetextcopyrightpermission[1]{}

\begin{abstract}
The majority of voice-based conversational agents still rely on pause-and-respond turn-taking, leaving interactions sounding stiff and robotic. We present RESPOND~(Responsive Engagement Strategy for Predictive Orchestration and Dialogue), a framework that brings two staples of human conversation to agents: timely backchannels~(“mm-hmm,” “right”) and proactive turn claims that can contribute relevant content before the speaker yields the conversational floor. Built on streaming ASR~(Automatic Speech Recognition) and incremental semantics, RESPOND continuously predicts both when and how to interject, enabling fluid, listener-aware dialogue. A defining feature is its designer-facing controllability: two orthogonal dials, Backchannel Intensity~(frequency of acknowledgments) and Turn Claim Aggressiveness~(depth and assertiveness of early contributions), can be tuned to match the etiquette of contexts ranging from rapid ideation to reflective counseling. By coupling predictive orchestration with explicit control, RESPOND offers a practical path toward conversational agents that adapt their conversational footprint to social expectations, advancing the design of more natural and engaging voice interfaces.
\end{abstract}



\keywords{Turn Prediction, Conversational Agents, Voice User Interfaces, Large Language Models, Human-Computer Interaction, User
Engagement}



\maketitle

\section{Introduction}
\label{sec:introduction}

Voice-based conversational agents have rapidly transitioned from novelty to ubiquity across smartphones, smart speakers, cars, and productivity platforms. Yet their interaction management still diverges from human dialogue: most commercial systems follow a rigid, half-duplex “pause-and-respond” protocol that waits for user silence before producing output. This approach simplifies engineering pipelines but often yields stilted, mechanical exchanges \cite{skantze2021turn}. In contrast, human conversations routinely exhibit sub-second timing, overlaps, and rapid response planning \cite{levinson2015timing}. Empirically, shorter response latencies are associated with higher conversational enjoyment, and overlapping interjections do not necessarily reduce perceived quality \cite{reece2023candor}. This gap between functional capability and conversational fluency remains a central HCI challenge.

Human conversation is not a sequence of neatly separated turns; it is a continuous, collaborative performance coordinated through subtle verbal and nonverbal cues. Timing, prosody, gaze, and micro-gestures regulate the flow of talk \cite{levinson2015timing}. Two mechanisms are especially important for maintaining fluidity: \emph{backchannels} and \emph{cooperative turn claims}.

\emph{Backchannels}~(e.g.,~“mm-hmm,” “right,” “I see”) are short listener acknowledgments produced while the speaker retains the conversational floor. They signal attention and understanding, and their timing is strongly cued by prosodic patterns \cite{ward2000prosodic}. When absent in agents, users often attribute the silence to computational delay or a lack of responsiveness. Recent predictive models treat backchannels as interaction opportunities rather than mere reactions, enabling real-time, frame-wise predictions of \emph{when} and even \emph{what} to produce \cite{inoue2024real,inoue2024yeah,lin-etal-2025-predicting}.

\emph{Cooperative turn claims} occur when the listener contributes after or before the speaker has formally yielded the turn, but in a supportive, collaborative manner rather than as a disruptive interruption. Conversation-analytic work shows that some “interruptions” function as cooperative overlap that advances common ground \cite{weatherall2018speakers}. In interactive systems, principled barge-in policies and interruption handling reduce latency while avoiding harmful cut-ins \cite{raux-eskenazi-2009-finite,zhao-etal-2015-incremental,khouzaimi2018methodology}. Moreover, user studies, especially with older adults, report higher perceived naturalness and engagement when agents allow user barge-in and offer timely acknowledgments during speech \cite{ding2022talktive,liu2025toward}.

In this work, we extend the notion of cooperative turn claims into a unified label, \textit{turn claim}, representing any listener attempt to claim the conversational floor, whether through interruption, overlap, or smooth turn exchange. This definition contrasts with \textit{backchannel}, which captures brief, supportive cues that signal engagement without seeking the floor, and \textit{stay silent}, where the listener provides no verbal response. Representative examples of these three behaviors are illustrated later in Section~\ref{sec:data}.

This paper introduces RESPOND~(\textit{Responsive Engagement Strategy for Predictive Orchestration and Dialogue}), a framework that bridges rigid machine turn-taking and the dynamic qualities of human conversation. RESPOND enables agents to generate both timely backchannels and cooperative turn claims, simulating a full-duplex architecture in which the system listens and speaks concurrently. Built on streaming ASR and incremental semantics, RESPOND continuously analyzes unfolding speech, and its \emph{Predictive Orchestration Engine} forecasts opportune moments for interjection, mirroring the fluidity of natural dialogue.

A key innovation lies not only in predictive capability but also in explicit \emph{controllability}. What counts as an appropriate interruption in brainstorming may be intrusive in counseling. RESPOND exposes a designer-facing control paradigm with two orthogonal dials:
\begin{itemize}
  \item \textbf{Backchannel Intensity}: tunes the frequency of acknowledgments, from a reserved listener to a highly engaged participant.
  \item \textbf{Turn Claim Aggressiveness}: modulates the propensity to contribute proactively, from a system that waits for turn completion to an assertive collaborator.
\end{itemize}
This level of control is achieved architecturally through lightweight conditioning~(e.g.,~feature-wise modulation), which allows for adjustments without retraining the core model \cite{perez2017filmvisualreasoninggeneral}. By making conversational behavior steerable and transparent, RESPOND aligns with intelligent user interfaces that prioritize user and designer agency.

This work makes the following contributions:
\begin{enumerate}
  \item We present RESPOND, which integrates predictive models for both backchanneling and cooperative turn claims within a real-time streaming pipeline.
  \item We introduce orthogonal dials for Backchannel Intensity and Turn Claim Aggressiveness, providing explicit, fine-grained control over an agent’s conversational style.
  \item We benchmark predictive accuracy against state-of-the-art turn-taking/backchanneling models and conduct a pilot study demonstrating the effect of tunable dials on perceived naturalness, responsiveness, and engagement.
\end{enumerate}
Together, these contributions enable conversational agents to respond in ways that feel more collaborative and human-like, bridging the gap between reactive systems and truly interactive dialogue partners.

\section{Related Work}

\subsection{Predictive turn-taking}
Classic systems determine speaking opportunities using voice activity detection~(VAD) and silence thresholds, often resulting in long pauses and delayed responses. In contrast, modern predictive turn-taking models~(PTTMs) aim to forecast both who will speak next and when, allowing agents to plan responses before the conversational floor is released. Voice Activity Projection~(VAP) learns to project joint future voice activity directly from audio within short~(e.g.,~2s) windows, supporting both turn-shift and backchannel inference \cite{ekstedt2022voice,ekstedt2022much}. Recent real-time implementations have demonstrated that VAP can operate continuously and efficiently on CPUs \cite{inoue2024real}. Multimodal extensions such as MM-VAP incorporate facial pose, gaze, and action units to improve hold/shift accuracy, particularly under overlapping speech \cite{russell-harte-2025-visual}. Further work has expanded predictive modeling to triadic settings \cite{lee2023multimodal, lee2024computational, elmers2025triadic}. Earlier decision-theoretic approaches also framed turn-taking—and even barge-in behavior—as a policy optimization problem \cite{raux-eskenazi-2009-finite,zhao-etal-2015-incremental}, while recent surveys synthesize key advances and remaining gaps in the field \cite{castillo-lopez-etal-2025-survey,ji2024wavchat}. Yet, existing PTTMs largely fall short in modeling more interactive behaviors such as interruption, overlapping speech, and backchanneling.

\subsection{Backchannel and turn claim prediction}
Backchannels~(e.g.,~“mm-hmm,” “yeah”) play a critical role in signaling attention and establishing shared understanding. Early efforts to predict backchannel and turn-taking behavior were typically unimodal. For instance, the VAP model \cite{ekstedt2022voice} mapped acoustic features into a 256-dimensional embedding to estimate the likelihood of turn-keeping, turn-shifting, and backchannel events. In the linguistic domain, Ekstedt and Skantze \cite{ekstedt2020turngpt} fine-tuned GPT-2 \cite{radford2019language} with explicit TURN tokens, achieving promising performance. More recent studies \cite{shukuri2023meta,kim2025beyond} highlight that large language models~(LLMs) are particularly effective at extracting contextual cues from text, often outperforming conventional approaches. Beyond speech and text, visual cues such as eyebrow and mouth movements have also been shown to shape turn-taking in face-to-face interactions \cite{lee2010predicting}.

Building on these foundations, multimodal approaches integrate complementary signals to improve robustness. Chang et al. \cite{chang2022turn} classified turn-taking into six sub-categories and introduced an end-to-end ASR-based framework that fused acoustic and linguistic features for both turn-taking and backchannel prediction. Yang et al. \cite{yang2022gated} further demonstrated that gated fusion of acoustic and semantic inputs improved accuracy, especially when textual cues augmented audio-only models. Other works \cite{kurata2023multimodal,wang2024turn} have similarly combined multimodal signals, achieving state-of-the-art results, while extensions to triadic interactions further highlight the benefits of cross-modal fusion \cite{lee2024online}.

Despite these advances, unimodal systems powered by the latest LLMs remain attractive for real-time deployment as they can achieve competitive accuracy while offering lower latency and reduced computational overhead compared to full multimodal pipelines.


\subsection{Proactive, barge-in–capable voice agents}
User studies consistently report that agents feel \emph{more natural} when they acknowledge, tolerate, and perform interactions appropriately in flight. For older adults, TalkTive showed that agent backchannels improve engagement during sensitive screening dialogs \cite{ding2022talktive}. Follow-ups that \emph{support both interruptions and backchannels} find higher perceived naturalness, fluency, and engagement than strict turn-by-turn baselines \cite{liu2025toward}. Beyond lab settings, meeting-scale work has found failed interruptions to improve inclusiveness in remote collaboration \cite{fu2022improving}. Linguistic and HCI studies highlight that “interruptions” are partly cultural and can be \emph{cooperative overlap}, not just dominance \cite{hilton2018does, hilton2016perception, weatherall2018speakers,cantrell2013power}. 

Beyond speech, overlap also benefits text-based agents. Allowing controlled “textual over-talk”~(parallelized messages) improves throughput and perceived flow, echoing speech literature on overlap as a cooperative device \cite{kim2025beyond}.

Our prototype integrates streaming ASR with incremental semantics extracted from an LLM, enabling the RESPOND module to determine in real time whether the agent should respond.

\section{Methodology}

We developed a text-based module for predicting listener behaviors such as backchannels and interruptions in real-time conversation. The system takes transcribed speech from a voice-based conversational agent and continuously outputs predictions for when the agent should backchannel, attempt to take the floor, or stay silent. Our design emphasizes lightweight modeling for low-latency deployment and controllability through scalar parameters that modulate the frequency of backchannels and the aggressiveness of interruptions.


\subsection{Data Preparation and Analysis}
\label{sec:data}
We use two text-based conversational corpora annotated for turn-taking behaviors: the MM-F2F dataset \cite{lin-etal-2025-predicting} and the CANDOR corpus \cite{reece2023candor}.

\begin{table}[ht]
\centering
\caption{Examples of input utterances and their corresponding labels from the MM-F2F dataset.}
\begin{tabular}{|c|c|}
\hline
\textbf{Input} & \textbf{Label} \\
\hline
last week yes we stayed home and we & KEEP \\
\hline
is hard to decide i just can't pick one & BACKCHANNEL \\
\hline
hi hi good afternoon how are you doing & TURN \\
\hline
\end{tabular}
\label{tab:mmf2fExample}
\end{table}

The MM-F2F dataset consists of face-to-face conversations with multimodal streams, from which we use only the transcripts. For the text-only task and for comparison with prior work, this dataset offers a train-ready format where incoming utterances are annotated with listener response types: \textit{TURN}, \textit{BACKCHANNEL}, and \textit{KEEP}. However, the paper overlooks interruption and overlap behaviors.  Table~\ref{tab:mmf2fExample} presents three illustrative examples, each corresponding to a different label.


The CANDOR corpus \citep{reece2023candor} comprises over 850 hours of video-call conversations with detailed annotations.  
Unlike MM-F2F, this dataset requires additional collation and preprocessing before it can be used for training.  
We first derive frame-level speaking status by aligning word-level transcript timestamps, and we adopt the Backbiter\footnote{Backbiter is a turn-model introduced in \citet{reece2023candor} that identifies short listener utterances, typically fewer than three words, composed largely of backchannel lexicon items, and not beginning with self-referential phrases (e.g., “I’m…”), as backchannels that occur in parallel with the speaker’s turn rather than as interruptions.} format to extract explicit backchannel instances as marked in the corpus.

The transcripts were segmented into 5-second windows with a 50~ms stride, yielding near–word-level inference.  
In rare cases of very rapid speech (e.g., “in the”), multiple words may fall within the same 50~ms interval.  
Each window was represented from two perspectives: $w^{A}$, where participant A is treated as the listener and participant B as the speaker, and $w^{B}$, where A and B switch roles.  
From each perspective, the model is trained to predict the appropriate class at the next word boundary given the listener’s viewpoint.  
This dual-perspective framing ensures that every conversation contributes two complementary sets of training data—one for A-as-listener and one for B-as-listener.  
Combined with the fine temporal resolution, it allows the model to capture rapid backchannels or short interruptions multiple times across overlapping windows, providing dense and balanced training coverage.

\begin{figure*}[ht]
    \centering
    \includegraphics[width=\linewidth]{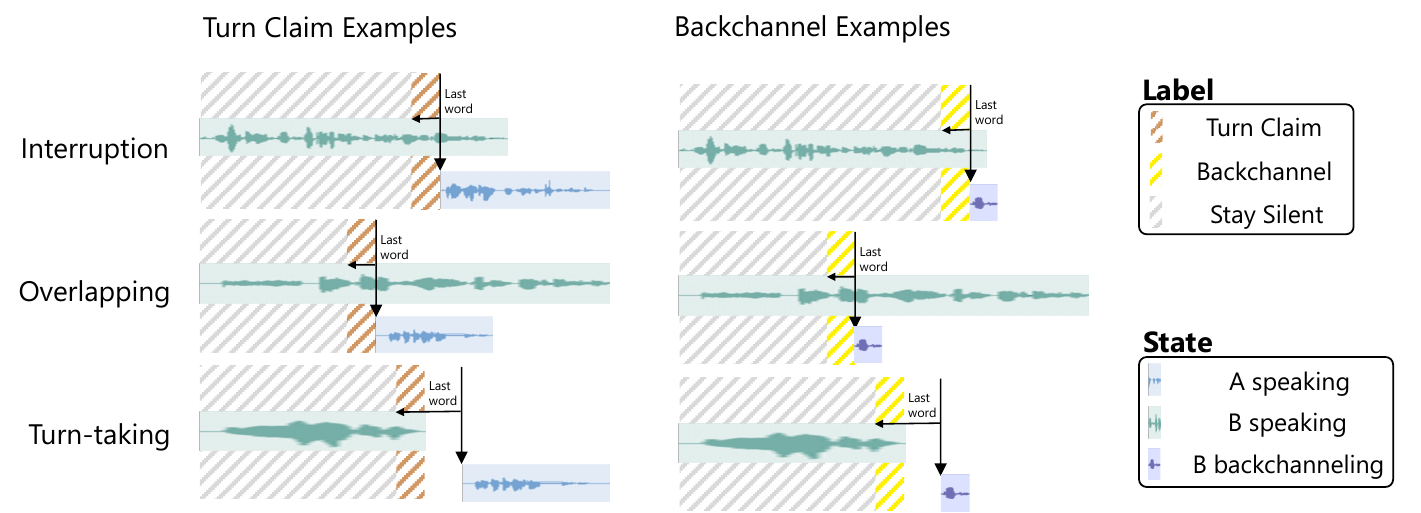}
    \caption{Illustration of the three annotation labels—\emph{turn claim}, \emph{backchannel}, and \emph{stay silent}—and how they appear in different conversational contexts. Left: listener actions categorized as turn claims (interruption, overlap, turn-taking). Right: corresponding backchannel examples occurring in similar contexts. Gray-shaded regions represent \emph{stay silent} intervals where the listener produces no response. Color bars indicate who is speaking or backchanneling.}
    \label{fig:label}
    \Description{All label situations}
\end{figure*}

Following the conceptual definition introduced in Section~\ref{sec:introduction}, we then operationalize three listener behaviors for modeling: \textit{turn claim}, \textit{backchannel}, and \textit{stay silent}.  
Figure~\ref{fig:label} illustrates representative situations for each category.  
We define \textit{turn claim} as any listener attempt to take the conversational floor, not necessarily resulting in a turn shift.  
Such attempts may occur through (i) \textit{interruption}, where the listener barges in and halts the speaker; (ii) \textit{overlapping speech}, where the listener speaks simultaneously with the speaker; or (iii) \textit{normal turn-taking}, where the listener takes the floor after the speaker finishes.  
In contrast, \textit{backchannel} refers to the same timing contexts, but the listener produces brief supportive cues (e.g., “uh-huh,” “yeah”) without attempting to claim the floor, following the annotation scheme of the corpus.  
Finally, any segment in which only the speaker talked—without interruption, overlap, or backchannel from the listener—was labeled as \textit{stay silent}.  
Each word boundary was annotated with one of these categories by aligning transcript content with the corresponding timing metadata.

\begin{figure*}[ht]
    \centering
    \includegraphics[width=\linewidth]{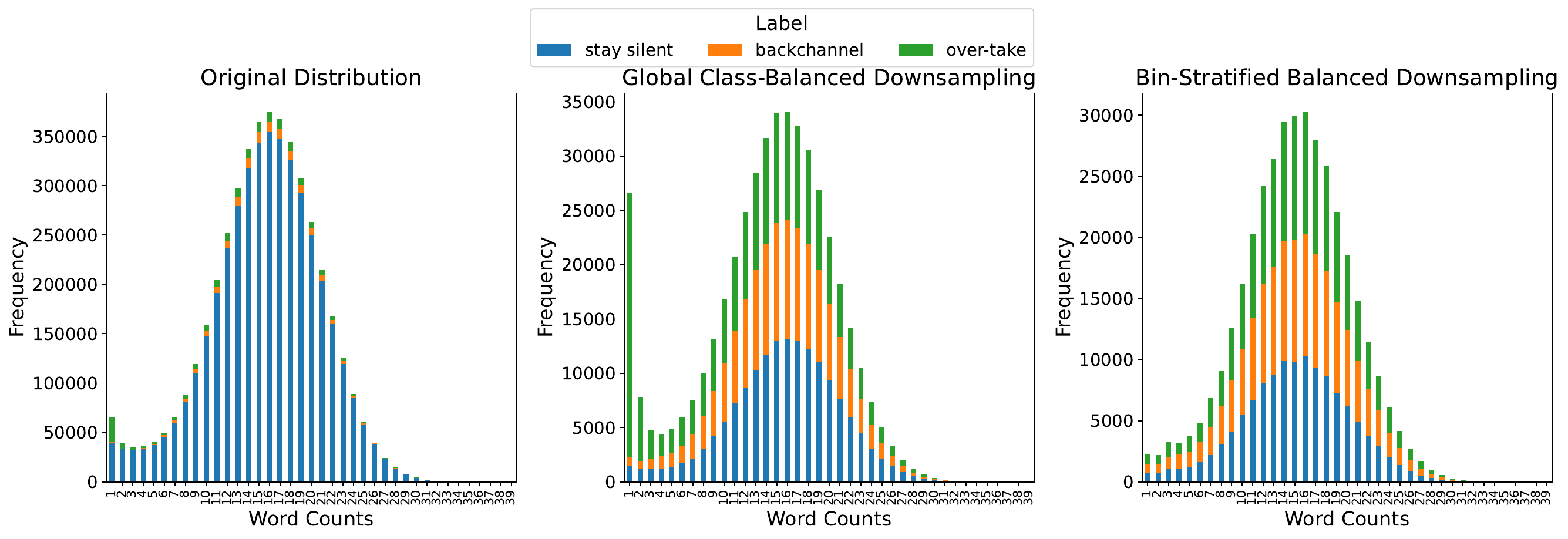}
    \caption{Label distributions before and after downsampling.}
    \label{fig:distribution}
    \Description{Showing how we deal with imbalanced data.}
\end{figure*}

To address class imbalance and reduce biases from variable word-count within a window, we applied bin-stratified balanced downsampling. Figure~\ref{fig:distribution} illustrates the effect of this procedure. The leftmost histogram shows the original distribution of window lengths, which is highly skewed. If we instead perform global class-balanced downsampling~(middle), label proportions are equalized overall, but short windows of one or two words remain dominated by the \textit{turn claim} class. This occurs because many backchannel windows consist of brief acknowledgments such as “yeah” or “interesting,” which are often followed by a speaker response and thus labeled as \textit{turn claim}. Our method~(right) alleviates this issue by balancing labels separately within each length bin.  

Concretely, for each window \(w_i\) we first computed its word-count $ C_i = \bigl|\text{split}(w_i)\bigr|$. We then assigned windows to predefined bins:  $\{1\}, \{2\},\; \{3\text{--}4\},\; \{5\text{--}6\},\; \ldots,\; \{29\text{--}30\},\; \{31+\}$. The first two bins are treated separately because single-word and two-word windows exhibit distinct dynamics: 
most backchannels fall within these lengths, while single-word windows are often dominated by \textit{turn claim} responses.
Grouping them together would therefore preserve a residual imbalance just like the global class-balanced downsampling in the middle of Figure~\ref{fig:distribution}. Within each bin \(b\), we considered all labels \(y \in \mathcal{Y}\)~(with \(\mathcal{Y} = \{\textit{turn claim}, \textit{backchannel}, \textit{stay silent}\}\)) and determined the smallest class count in that bin:
\begin{equation}
    n_b = \min_{y \in \mathcal{Y}} \; \#\{\,~(w_i, y_i) \mid C_i \in b,\; y_i = y \,\}
\label{equa:smallest}
\end{equation}

We then randomly sampled \(n_b\) windows for each label, ensuring that every \((\text{bin}, \text{label})\) bucket contained the same number of examples. Before downsampling, the dataset contained 132{,}277 \textit{turn claim}, 159{,}787 \textit{backchannel}, and 4{,}277{,}563 \textit{stay silent} windows, showing a strong class imbalance. After applying the bin-stratified balanced downsampling, each class was reduced to 127{,}674 samples, resulting in a balanced dataset. The downsampled sets were then concatenated, shuffled, and split into training, validation, and test subsets with a ratio of 18:1:1.

\subsection{Model}
The backbone of our system is the Qwen3-0.6B language model \cite{qwen2025qwen3technicalreport}, a 0.6B-parameter Transformer designed for efficiency. We fine-tune the model using Low-Rank Adaptation~(LoRA~\cite{hu2021loralowrankadaptationlarge}) to adapt it for turn-taking prediction while keeping the parameter footprint manageable. The model outputs a classification over three categories: turn claim, backchannel, or stay silent. 

\subsection{Controllability}
\label{sec:control}

A key goal of our framework is to allow flexible control over the conversational style of the agent. To this end, we introduce two continuous control parameters: \textit{backchannel intensity}~($c_{\text{bc}}$) and \textit{turn claim aggressiveness}~($c_{\text{tc}}$). These values capture how frequently a listener tends to produce backchannels or take the floor during a conversation. By conditioning the model on these values, we can bias predictions toward more passive or more assertive behaviors without retraining the underlying network.

\paragraph{\textbf{Calculation of control parameters.}}
The values for the control parameters are computed at the conversation level for each participant. In a dyadic conversation, this yields two distinct sets of control parameters, one for participant A and one for participant B.  

Suppose a conversation consists of $N$ total frames. Let $N_{\text{bc}}^{A}$ and $N_{\text{spk}}^{A}$ denote the number of frames in which participant~A is backchanneling and speaking, respectively, and let $N_{\text{bc}}^{B}$ and $N_{\text{spk}}^{B}$ denote the corresponding counts for participant B. The control parameters are then defined as:
\begin{displaymath} 
c_{\text{bc}}^{A} = \frac{N_{\text{bc}}^{A}}{N}, 
\qquad
c_{\text{tc}}^{A} = \frac{N_{\text{spk}}^{A}}{N},
\qquad
c_{\text{bc}}^{B} = \frac{N_{\text{bc}}^{B}}{N},
\qquad
c_{\text{tc}}^{B} = \frac{N_{\text{spk}}^{B}}{N}.
\end{displaymath}

Here, $c_{\text{bc}}$ reflects the proportion of time a participant produces backchannels, while $c_{\text{tc}}$ represents their overall turn claim ratio—that is, the proportion of frames in which the participant speaks (voiced activity excluding backchannels), regardless of whether they currently hold the conversational floor.  
These counts are computed over the entire conversation, capturing each participant’s general speaking style rather than properties of individual windows.  
When extracting training segments, each window inherits the control values of the designated listener.  
As a result, the same conversation yields two complementary sets of windows:  one where A is modeled as the listener~(with $(c_{\text{bc}}^{A}, c_{\text{tc}}^{A})$), and another where B is modeled as the listener~(with $(c_{\text{bc}}^{B}, c_{\text{tc}}^{B})$).  
This ensures that control parameters remain consistent with the overall behavioral tendencies of the selected listener.

\begin{figure*}[ht]
    \centering
    \includegraphics[width=\linewidth]{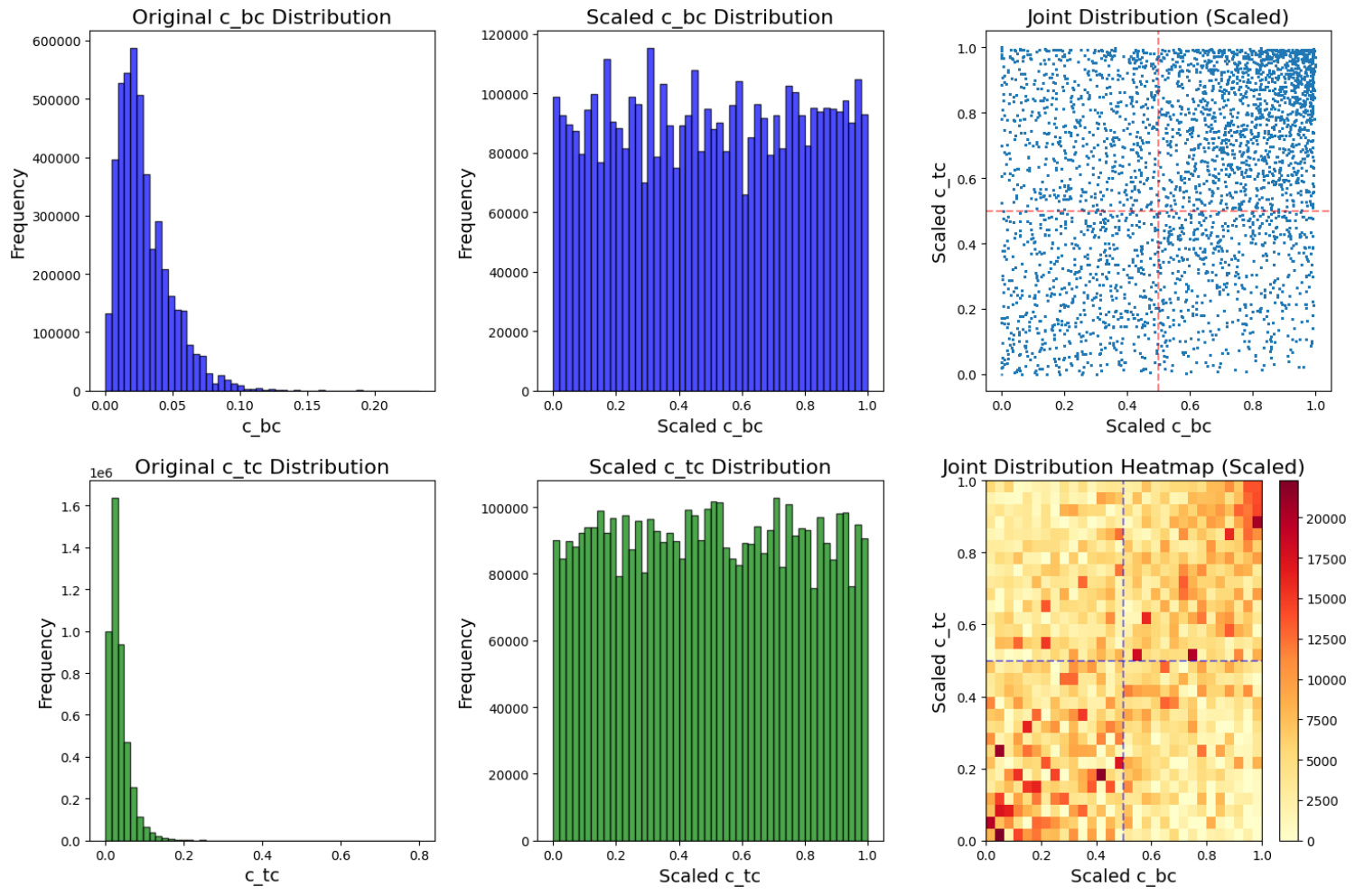}
    \caption{Mapping control parameters to a uniform scale. Here, $c_{\text{bc}}$ denotes backchannel intensity and $c_{\text{tc}}$ denotes turn claim aggressiveness. The left column shows the raw distributions of the calculated ratios, the middle column shows the distributions after quantile-based transformation, and the right column visualizes the scaled values as both scatter points and a heatmap.}
    \label{fig:control_scale}
    \Description{Shows how we scale the control values and they are uniformly distributed.}
\end{figure*}

Instead of using the raw control parameters directly, we normalize them with a quantile-based transformation to distribute the values more evenly over the range $[0,1]$. Specifically, we apply the \texttt{QuantileTransformer} from \texttt{scikit-learn}~\cite{scikit-learn}, which maps the empirical distribution of the data to a uniform distribution such that each quantile is equally represented. Figure~\ref{fig:control_scale} illustrates the distributions before and after transformation. This step addresses the skewed nature of the raw ratios: for example, many conversations have backchannel intensities $c_{\text{bc}}$ concentrated below $0.15$. Without rescaling, most values would lie in a narrow band near zero, making it difficult for the model to learn meaningful distinctions. By contrast, the quantile transformation spreads these differences across the full $[0,1]$ range, improving training stability. In addition, the rescaled values provide a more intuitive interface for controllability at inference time, allowing users to interpret settings on a normalized 0–100\% scale rather than in dataset-specific raw proportions.  



\paragraph{\textbf{Integration into the model.}}
We condition the model on these control values using Feature-wise Linear Modulation~(FiLM) layers \citep{perez2017filmvisualreasoninggeneral,dumoulin2018feature-wise}. Given a hidden representation $h \in \mathbb{R}^d$ from the backbone encoder, we compute:
\[
\text{FiLM}(h; c) = \gamma(c) \odot h + \beta(c),
\]
where $c = [c_{\text{bc}}, c_{\text{tc}}]$ and $\gamma(\cdot), \beta(\cdot)$ are small multi-layer perceptrons that transform the control scalars into per-dimension scaling and shifting factors. This operation adjusts the hidden activations in a way that reflects the overall interaction style of the conversation, without modifying the pretrained backbone. During training, each input window is paired with the $(c_{\text{bc}}, c_{\text{tc}})$ values computed from its parent conversation. This ensures that the model learns not only local turn-taking dynamics, but also how these dynamics should adapt under different global styles. For example, in a conversation with a high $c_{\text{bc}}$, the model sees all windows conditioned on a “frequent backchannel” setting, encouraging predictions that align with that style.

\begin{figure*}
    \centering
    \includegraphics[width=\linewidth]{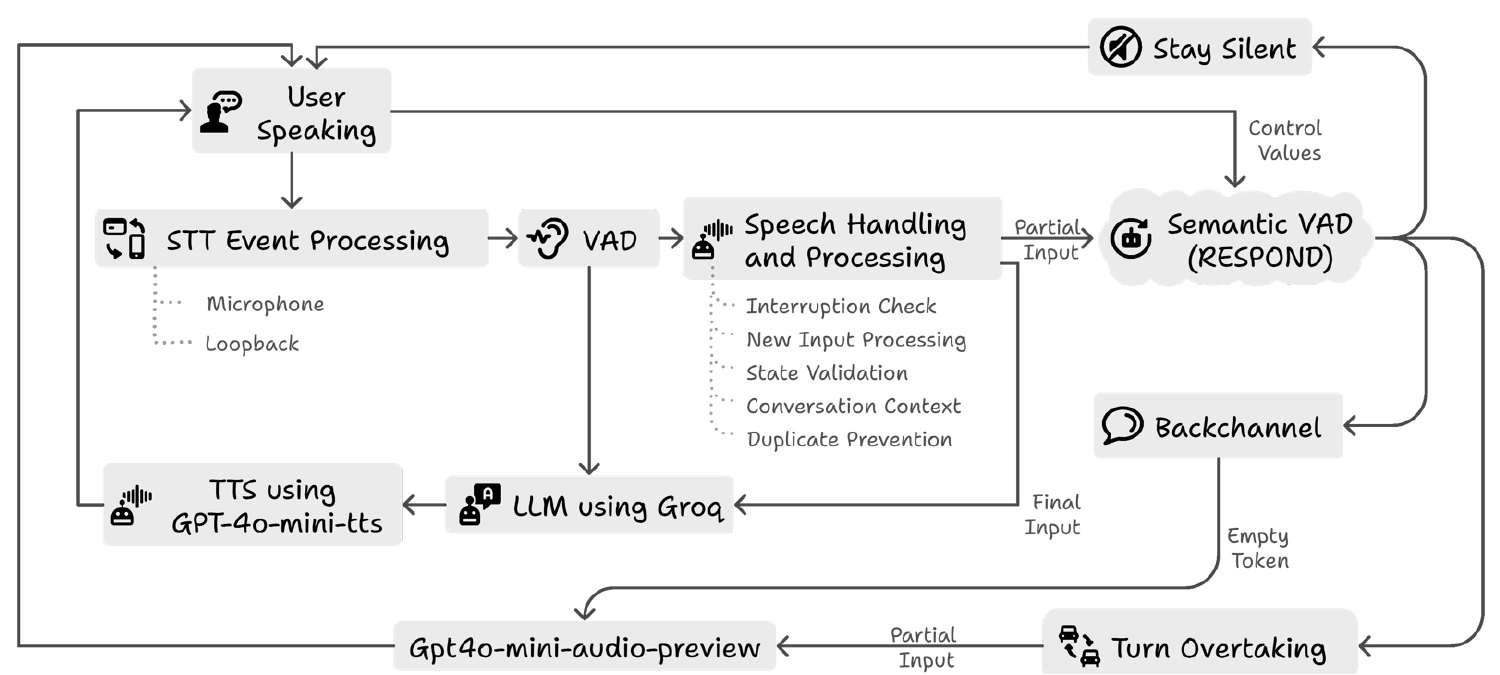}
    \caption{System pipeline of our conversational agent module.}
    \label{fig:pipeline}
    \Description{System logic and flow.}
\end{figure*}

\subsection{System Design}
Figure~\ref{fig:pipeline} illustrates the end-to-end inference pipeline of RESPOND. Incoming speech is first transcribed into word-level transcripts by an automatic speech recognition~(ASR) module. The ASR is powered by Microsoft Azure Speech Cognitive Services.\footnote{\url{https://speech.microsoft.com/portal}} The transcripts are accumulated into sliding windows of up to 5 seconds for inference.  

This pipeline supports real-time backchannel and turn claim prediction based on the most recent partial transcript.  

Although inference itself operates at word-level granularity, overall responsiveness is constrained by ASR latency—the time between a spoken word and its textual availability.  
In our implementation, this delay typically ranges between 250–500~ms under stable network conditions, consistent with modern streaming ASR benchmarks.  
Consequently, the timing of generated backchannels may be offset by this recognition latency, an inherent limitation of text-based real-time systems.

\textit{RESPOND} module provides user-level customization of interaction style by exposing controllable parameters. In particular, users can specify $(c_{\text{bc}}, c_{\text{tc}})$ to adjust the agent’s communication style: higher $c_{\text{bc}}$ encourages more frequent backchannels, while higher $c_{\text{tc}}$ increases the likelihood of overtaking the conversational floor:

\begin{itemize}
    \item \textbf{Passive listener:} $c_{\text{bc}}=0.1$, $c_{\text{tc}}=0.0$~(rare backchannels, never interrupts).  
    \item \textbf{Collaborative listener:} $c_{\text{bc}}=0.6$, $c_{\text{tc}}=0.2$~(frequent backchannels, occasional turn-taking).  
    \item \textbf{Assertive speaker:} $c_{\text{bc}}=0.1$, $c_{\text{tc}}=0.8$~(infrequent backchannels, frequent interruptions or floor-taking).  
\end{itemize}

Alternatively, the system can estimate $c_{\text{bc}}$ and $c_{\text{tc}}$ dynamically from past conversation history, allowing the agent to mirror the style of its partner. In either case, controllability enables fine-grained adaptation of behavior with negligible computational cost. Then, the resulting decision is passed to a text-to-speech~(TTS) system that generates the corresponding verbal response. This architecture ensures that the agent can provide timely acknowledgments or strategically take the floor, creating a more natural interaction flow.

\subsection{Experiments}

\paragraph{\textbf{Implementation Details}}
All experiments were implemented in PyTorch and trained on a single NVIDIA A100 GPU. We used the Qwen3-0.6B language model as the backbone, which has a hidden size of 768. For controllability, we injected two scalar values, backchannel intensity $c_{\text{bc}}$ and turn claim aggressiveness $c_{\text{tc}}$, through a FiLM layer. Each FiLM layer consisted of two one-layer MLPs with ReLU activations, which projected the control scalars into scaling and shifting vectors of size 768 to match the backbone hidden states.  

Models were fine-tuned using LoRA with rank $r=16$ and $\alpha=32$, enabling efficient adaptation with a limited parameter budget.  Training was performed with a per-device batch size of 128 for both training and evaluation. We optimized using AdamW with a learning rate of $1\times 10^{-5}$, weight decay of 0.01, cosine learning rate scheduling, and a warm-up ratio of 0.1. Gradient clipping was applied with a maximum norm of 1.0. We monitored validation loss at the end of each epoch and selected the checkpoint with the lowest loss as the final model. Training was terminated after three epochs, as validation performance stabilized and earlier checkpoints showed no further improvement. These settings provided stable convergence and good generalization while keeping computational time feasible for rapid iteration. A complete training run required approximately 1.5 hours on the A100 GPU. We fixed the random seed to 42 and disabled data loader multiprocessing to ensure reproducibility.

\subsection{Pilot Study}
\begin{figure*}[ht]
    \centering
    \includegraphics[width=0.75\linewidth]{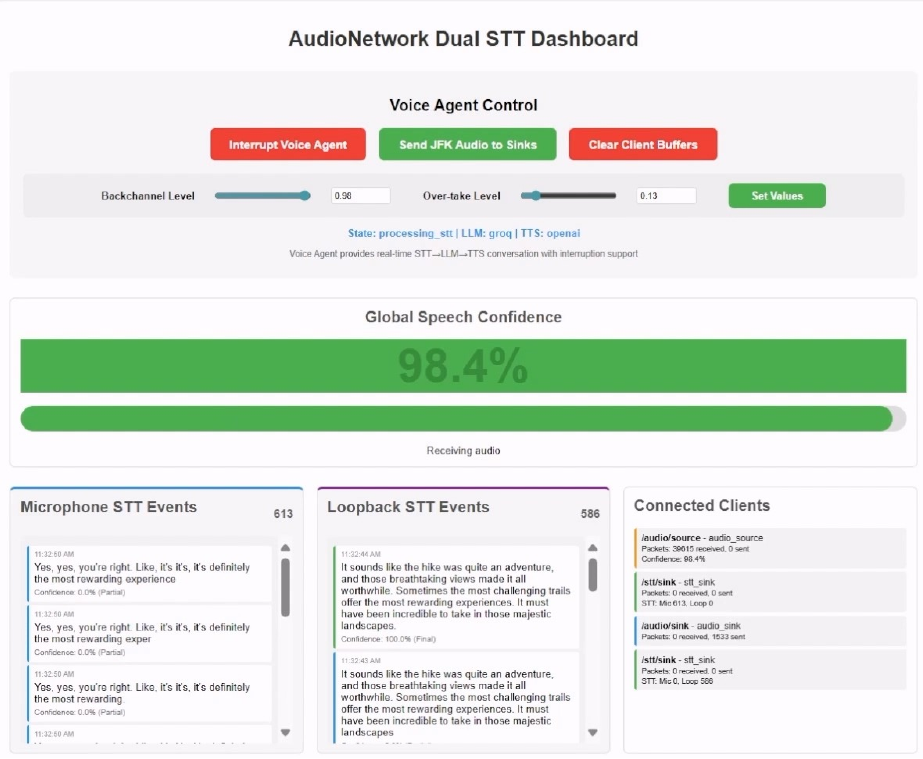}
    \caption{The user interface of our system.}
    \label{fig:demo}
    \Description{UI}
\end{figure*}

To complement the quantitative evaluation, we conducted a small-scale exploratory pilot study to examine how users perceived and interacted with the controllability features of our conversational agent. Six participants~(4 female, 2 male) were recruited from our lab community; all had prior experience with commercial voice assistants but no prior exposure to our system. Figure~\ref{fig:demo} shows the user interface, which provides two sliders that allow users to adjust backchannel intensity and turn claim aggressiveness on a continuous scale from 0 to 1.

\paragraph{\textbf{Procedure}} The study was reviewed and approved by the appropriate ethics review process at the authors’ institution and was determined to involve minimal risk. No personally identifying data were collected. Each participant first engaged in a 10–12 minute open-ended conversation with the system under a default configuration. Participants were instructed to interact naturally, as they would with a voice-based assistant, by asking questions, sharing opinions, or describing daily experiences.  
The goal was to elicit spontaneous, mixed-initiative exchanges rather than task-specific or scripted interactions. After this initial interaction, participants were introduced to the controllability interface, which exposed the two scalar parameters $(c_{\text{bc}}, c_{\text{tc}})$. They were encouraged to experiment with different settings by adjusting the sliders and immediately observing the changes in the system’s behavior. Once they were comfortable, participants were asked to continue conversing with the agent while iteratively tuning the values until they arrived at what they considered their “ideal” configuration. After the session, participants completed a short semi-structured interview focusing on naturalness, responsiveness, and overall satisfaction with both the default and self-selected settings.  

\section{Results}

\subsection{Benchmarking Against Prior Work}

We compare RESPOND against previously reported baselines from the MM-F2F benchmark \citep{lin-etal-2025-predicting}, which re-implemented and evaluated several representative models on the same dataset:  
(1) TurnGPT \citep{ekstedt2020turngpt}, which processes only text and predicts turn-hold versus turn-shift actions;  
(2) the model of \citeauthor{wang2024turn} \citeyearpar{wang2024turn}, which integrates text and audio features for joint turn-taking and backchannel prediction; and  
(3) the multimodal model of \citeauthor{kurata2023multimodal} \citeyearpar{kurata2023multimodal}, which fuses text, audio, and video streams for turn-taking prediction.  
For completeness, we also include the official MM-F2F baseline results (both multimodal and text-only configurations) reported by \citet{lin-etal-2025-predicting}, and add our own results obtained using the same data splits and evaluation protocol.

As shown in Table~\ref{tab:comparison-study}, RESPOND outperforms the text-only approaches, achieving higher F1-scores on the \textit{Keep} and \textit{Turn} classes as well as improved overall accuracy compared to the strongest text-only baseline. Although our framework does not surpass the latest multimodal system that combines text, audio, and video, it achieves competitive performance while being substantially more lightweight and efficient—making it suitable for real-time deployment.

All MM-F2F results are reported on the dataset’s native label distribution, without additional class rebalancing.  
Control parameters for backchannel intensity ($c_{\text{bc}}$) and turn claim aggressiveness ($c_{\text{tc}}$) are not applied in this experiment, as the MM-F2F corpus does not provide annotations for interruptions or overlapping speech.  
These controllable parameters are introduced and analyzed later using the CANDOR corpus, which explicitly includes such phenomena.




\begin{table}[ht]
\centering
\caption{The comparison study results on the MM-F2F dataset. We compare TurnGPT, \citeauthor{wang2024turn}, \citeauthor{kurata2023multimodal}, and the work from MM-F2F~\cite{lin-etal-2025-predicting} itself to our model. The \textbf{T}, \textbf{A}, \textbf{V} denote input modalities of text, audio, and video.}
\resizebox{\columnwidth}{!}{
\begin{tabular}{@{}l|l|c|ccc@{}}
\toprule\toprule
\textbf{Method} & \textbf{Modal} & \textbf{Acc.} & \multicolumn{3}{c}{\textbf{F1-Score}} \\
                &                &                   &\textbf{Keep} & \textbf{Turn} & \textbf{BC} \\
\midrule
TurnGPT~\cite{ekstedt2020turngpt}                            & T     & 0.645 & 0.745 & 0.420 & -     \\
\citeauthor{wang2024turn}'s         & T+A   & 0.737 & 0.742 & 0.739 & 0.680 \\
\citeauthor{kurata2023multimodal}'s & T+A+V & 0.720 & 0.729 & 0.728 & 0.667 \\
MM-F2F~\cite{lin-etal-2025-predicting}                                & T+A+V & 0.823 & 0.806 & 0.811 & 0.906 \\ 
MM-F2F                                & T & 0.751 & 0.747 & 0.767 & 0.707 \\ 
Ours                                & T & 0.756 & 0.770 & 0.773 & 0.697 \\ 
\bottomrule\bottomrule
\end{tabular}
}
\label{tab:comparison-study}
\end{table}

\subsection{Performance on the CANDOR Corpus}

We further evaluated our framework on the CANDOR corpus \citep{reece2023candor}, which comprises over 850 hours of video-call conversations annotated with detailed turn-taking behaviors.  
Unlike MM-F2F, which focuses on clean two-party exchanges, CANDOR captures a richer range of spontaneous conversational phenomena, including interruptions, overlaps, and frequent backchanneling, making it a challenging and comprehensive benchmark for testing both prediction accuracy and controllability.

For this experiment, we used the balanced training split provided by CANDOR to ensure equal representation of \emph{turn claim}, \emph{backchannel}, and \emph{stay-silent} behaviors.  
We also incorporated the controllability parameters introduced in Section~\ref{sec:control}:  
$c_{\text{bc}} \in [0,1]$ for \textit{Backchannel Intensity} and  
$c_{\text{tc}} \in [0,1]$ for \textit{Turn Claim Aggressiveness}.  
These scalar values were concatenated with the model’s hidden representation through FiLM conditioning, allowing fine-grained modulation of listener behavior during both training and inference.

\begin{figure}[ht]
    \centering
    \includegraphics[width=\linewidth]{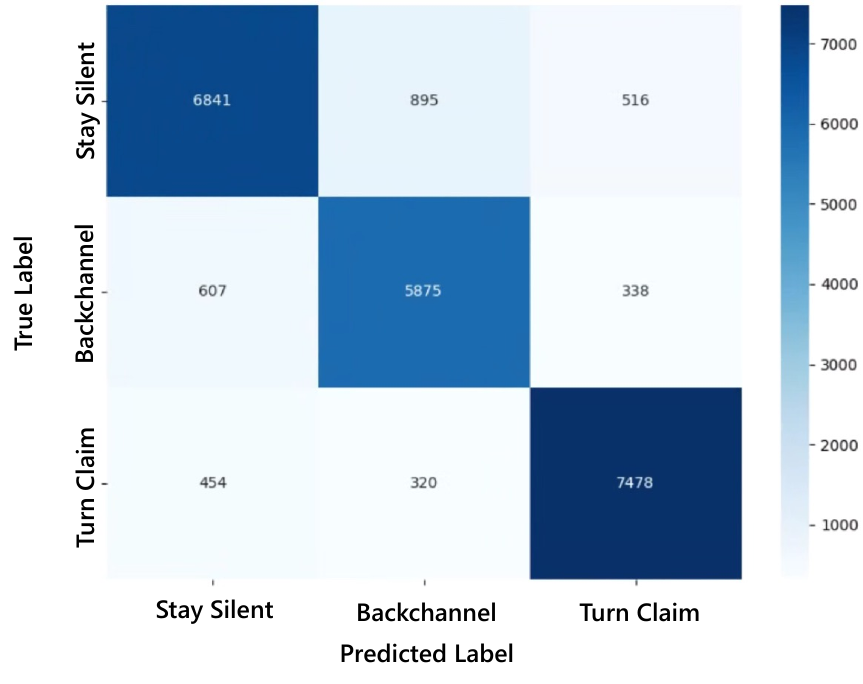}
    \captionof{figure}{Confusion matrix on the CANDOR test set.}
    \label{fig:confusion}
    \Description{confusion matrix of the result}
\end{figure}

Table~\ref{tab:candor_results} summarizes the model’s performance on the CANDOR test set.  
The model achieved F1-scores of 0.847 for \emph{turn claim}, 0.845 for \emph{backchannel}, and 0.902 for \emph{stay-silent}, with an overall accuracy of 0.87. Figure~\ref{fig:confusion} shows the confusion matrix, illustrating that most misclassifications occur between \emph{stay silent} and \emph{backchannel}. This confusion likely arises from conversational contexts in which a brief backchannel could be appropriate but remaining silent is also socially acceptable.

We report only our model’s results on this corpus, as existing baselines such as TurnGPT or the MM-F2F implementations were not designed to handle interruption or overlap categories and lack controllable inputs for backchannel and turn claim behaviors.  
Moreover, these models were originally optimized for dyadic turn-shift prediction under fixed label schemes, making direct retraining or comparison on CANDOR non-trivial and outside the scope of our controllability study.  
The purpose of this evaluation is therefore to assess the generalization and tunability of our controllable framework in a complex, naturalistic conversational setting.



\begin{table}[ht]
\centering
\caption{Performance of our model on the CANDOR corpus.}
\begin{tabular}{lccc}
    \toprule
    \textbf{Category} & \textbf{Precision} & \textbf{Recall} & \textbf{F1} \\
    \midrule
    Turn claim      & 0.8657 & 0.8290 & 0.8470 \\
    Backchannel     & 0.8286 & 0.8614 & 0.8447 \\
    Stay silent     & 0.8975 & 0.9062 & 0.9018 \\
    \midrule
    \textbf{Overall Accuracy} & \multicolumn{3}{c}{0.87} \\
    \bottomrule
\end{tabular}
\label{tab:candor_results}
\end{table}

\subsection{Case Studies}

\begin{figure*}[t]
    \centering
    \begin{subfigure}{0.9\linewidth}
        \centering
        \includegraphics[width=\linewidth]{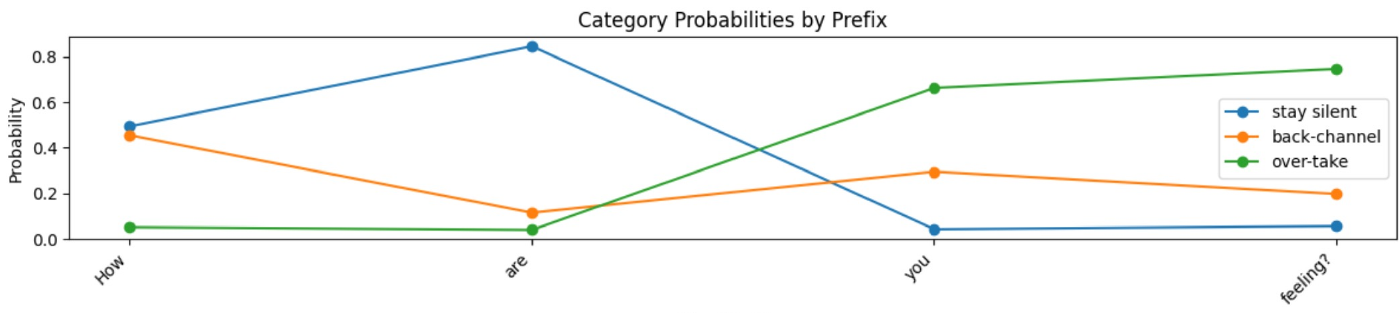}
        \caption{}
        \label{fig:case1}
    \end{subfigure}

    \begin{subfigure}{0.9\linewidth}
        \centering
        \includegraphics[width=\linewidth]{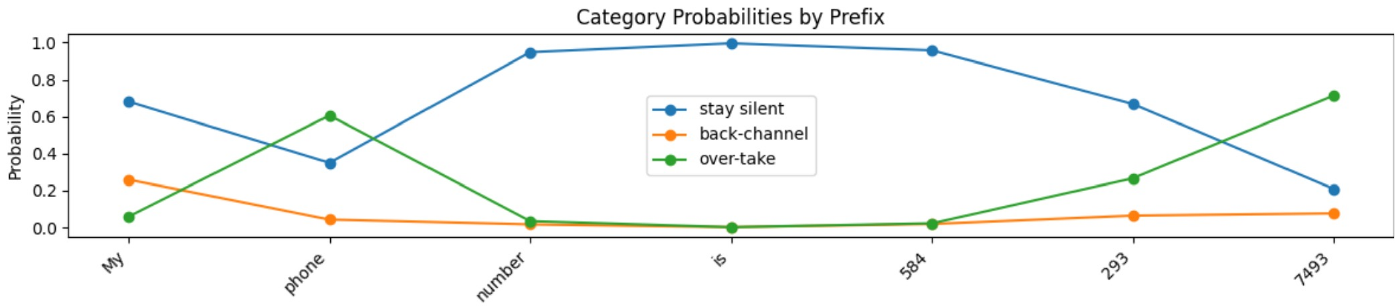}
        \caption{}
        \label{fig:case2}
    \end{subfigure}
    
    \begin{subfigure}{0.9\linewidth}
        \centering
        \includegraphics[width=\linewidth]{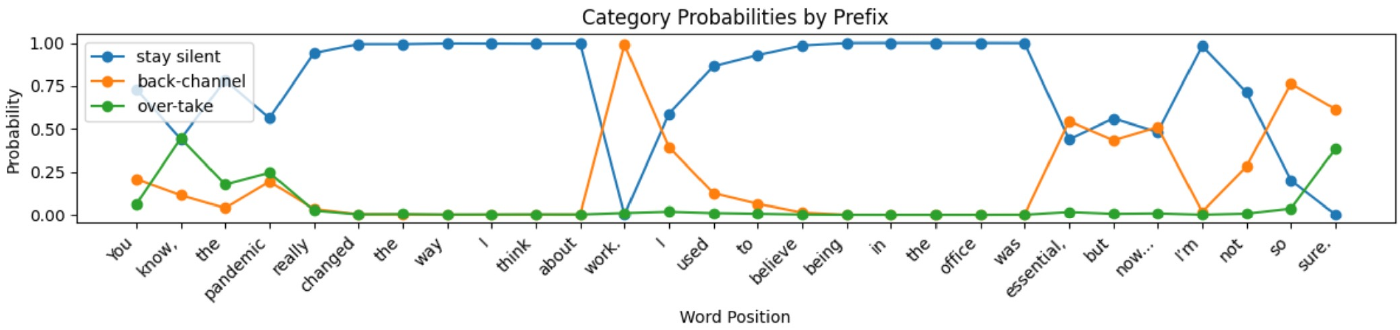}
        \caption{}
        \label{fig:case3}
    \end{subfigure}
    
    \caption{Illustrative examples of RESPOND predictions across three scenarios: a simple greeting, a task-oriented exchange, and a longer casual conversation.}
    \label{fig:cases}
    \Description{Show word by word prediction for different cases.}
\end{figure*}

In addition to quantitative evaluation with accuracy and F1-scores, we conducted a qualitative analysis to better understand the model’s real-time behavior.  
For this purpose, we randomly selected representative conversation segments, ranging from simple greetings to task-oriented and free-form interactions, and visualized how prediction probabilities evolved as words were incrementally appended to the observation window.  
These case studies reveal how the model differentiates among \emph{backchannel}, \emph{turn claim}, and \emph{stay-silent} behaviors, illustrating its sensitivity to conversational context and timing.

Figure~\ref{fig:case1} presents a simple greeting. The model predicts a turn claim slightly early right after the word “you,” though this can be considered reasonable since conversational overlaps often occur immediately after phrases such as “how are you.” Figure~\ref{fig:case2} illustrates a task-oriented interaction where the speaker provides a phone number. Although the model incorrectly predicts a turn claim at the second word, it successfully identifies the natural boundary at the end of the phone number and appears to capture the expected length of a formal phone number. Figure~\ref{fig:case3} shows a longer, casual conversation with hesitations. In this example, the model produces timely backchannels that encourage the speaker to continue, demonstrating context-sensitive responsiveness.

\subsection{User Experience}

\paragraph{\textbf{Findings}} Participants unanimously reported that adjusting the controllability parameters led to noticeable and meaningful differences in the system’s conversational style. Higher $c_{\text{tc}}$ values were described as making the agent “too pushy” or “constantly jumping in,” while lower $c_{\text{tc}}$ produced a calmer, more comfortable interaction. Backchannel intensity~($c_{\text{bc}}$) was generally well received when moderate, with participants noting that it “felt like someone was listening” or “kept the conversation flowing.” However, excessive backchannels, especially mid-sentence ones, were considered distracting and unnatural.  

\begin{figure*}[ht]
    \centering
    \includegraphics[width=0.33\linewidth]{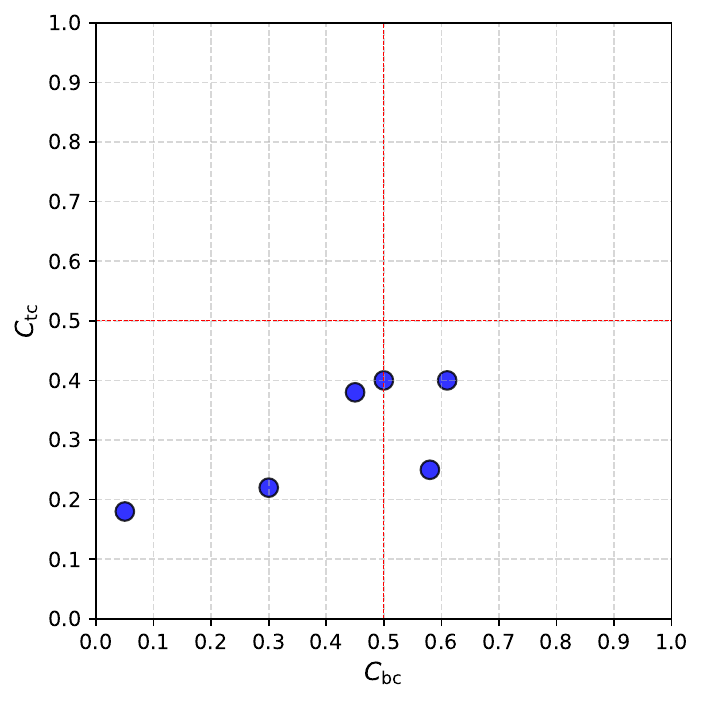}
    \caption{Distribution of user-selected control parameters.}
    \label{fig:user_control}
    \Description{user control parameters distribution}
\end{figure*}

\paragraph{\textbf{Preferences}} Figure~\ref{fig:user_control} shows the distribution of user-selected control parameters. Three out of six participants ultimately set $c_{\text{tc}}$ close to 0.2, indicating a preference for minimal turn claim. At the same time, most participants selected a moderate $c_{\text{bc}}$, suggesting that occasional, well-timed backchannels are desirable. Two participants opted for slightly higher backchannel intensities, stating that it made the system feel more engaged, but even they avoided high overtaking aggressiveness. Across the group, participants emphasized that the “ideal” configuration balanced attentiveness with restraint.  

From both the conversations and the interviews, several key observations emerged:  
\begin{itemize}
    \item \textbf{Timing matters:} sentence-final backchannels felt natural, but mid-sentence backchannels were disruptive.  
    \item \textbf{Overtaking caution:} frequent or mistimed turn claims eroded trust and willingness to engage.  
    \item \textbf{Perceived control:} participants appreciated being able to adjust values themselves, describing the sliders as “knobs” to personalize the system’s personality.  
    \item \textbf{Exploration benefit:} letting users experiment helped them understand the range of behaviors and settle on comfortable defaults.  
\end{itemize}

Overall, the study confirmed that the controllability framework produces clear, perceivable effects on conversational style, and that users prefer to operate at low overtaking aggressiveness with moderate backchannel intensity. The ability to interactively adjust these settings was particularly valued, as it gave participants a sense of agency and personalization. These findings suggest that controllability is not only technically effective but also user-friendly, though future work is needed to refine backchannel timing and reduce overly aggressive floor-taking.

\section{Discussion}
\label{sec:discussion}

RESPOND operationalizes two long-standing desiderata for conversational agents, timely backchannels and cooperative, context-appropriate turn claims, through a \textit{text-only, low-latency} module with \textit{explicit behavioral control}. Across MM-F2F and CANDOR, a lightweight unimodal approach delivers competitive accuracy while exposing interpretable dials that practitioners can tune to the social norms of a setting~(e.g.,~ideation vs.\ counseling). On MM-F2F, we show parity or modest gains versus text-only baselines while remaining far simpler than multimodal stacks; on CANDOR, the model maintains balanced performance across turn claim, backchannel, and stay-silent despite naturalistic overlaps and interruptions.

Our exploratory pilot user study suggests that this steerability is practically useful. Most participants preferred low turn claim aggressiveness~($c_{\text{tc}}\approx 0.2 - 0.4$) with moderate backchannel intensity~($c_{\text{bc}}\approx 0.4 - 0.6$), highlighting the importance of sentence-final timing for acknowledgments. A key design implication is that systems should default to conservative settings while allowing for user-specific adjustments.

RESPOND complements emerging ASR–LLM–TTS pipelines that already excel at streaming recognition and generation. Within these pipelines, RESPOND provides policy-level guidance on when and how to interject. Its FiLM-based conditioning integrates directly with a text encoder and can be updated online without retraining. Moreover, the system exposes quantile-scaled control parameters~($c_{\text{bc}}$, $c_{\text{tc}}$), which act as intuitive, human-interpretable controls, the same UI “knobs” that end users adjusted in our pilot study, highlighting a feasible path toward personalized, etiquette-aware conversational agents in production.

\section{Limitations and Future Work}
\label{sec:limitations}

Our current model is text-only; while this design maximizes deployability and minimizes latency, it omits prosodic and visual cues that are well known to signal backchannels and turn transitions. Future work may consider extending the model with prosody-aware features~(e.g.,~pause length, pitch movement) and lightweight visual signals~(e.g.,~head nods, eyebrow raises), incorporated through late fusion to preserve low runtime costs. For training, we currently collapse normal turn-taking, interruption, and overlap into a single turn claim class, simplifying dataset unification but masking socially important distinctions. A promising direction is to separate these subtypes and examine how controllability dials differentially modulate them—for instance, encouraging cooperative overlap while suppressing disruptive interruption.

Control parameters~($c_{\text{bc}}$, $c_{\text{tc}}$) are currently estimated from global ratios per participant; although quantile mapping mitigates skew, these estimates remain dataset-specific. Future work may consider developing online estimators that adapt to each conversation and introduce context-conditioned control~(task, relationship, culture) to support etiquette transfer across domains. Our pilot study~(N=6) surfaced useful patterns but was underpowered and used a fixed default$\rightarrow$tuning order; scaling to a larger, counterbalanced study with standardized measures~(naturalness, responsiveness, trust) and system-level metrics~(latency, barge-in error) will provide more robust evaluation. Finally, although we outline a real-time pipeline (ASR$\rightarrow$RESPOND$\rightarrow$TTS), comprehensive end-to-end duplex evaluation with overlapping streaming TTS remains future work.
Integration with open duplex frameworks~(e.g.,~TEN, Hertz-dev) and the addition of policy constraints~(role- and context-aware caps on $c_{\text{tc}}$) and explanatory UIs that make behavioral settings transparent are critical next steps toward safe and responsible deployment.





\bibliographystyle{ACM-Reference-Format}
\bibliography{sample-base}

\appendix


\end{document}